\documentclass[preprint2]{aastex}

\shorttitle{Globular Cluster Catalog}
\shortauthors{Harris}

\begin{document}

\title{A New Catalog of Globular Clusters in the Milky Way}

\author{William E. Harris}
\affil{Department of Physics \& Astronomy, McMaster University,
  Hamilton ON L8S 4M1}
\email{harris@physics.mcmaster.ca}

\begin{abstract}
A new revision of the McMaster catalog of Milky Way globular clusters
is available. This is the first update since 2003 and the biggest single revision since
the original version of the catalog published in 1996.  
The list now contains a total of 157 objects classified as globular clusters.
Major upgrades have been
made especially to the cluster coordinates, metallicities, and structural
profile parameters, and the list of parameters now also includes central
velocity dispersion. 

\emph{NB:}  This paper is a stand-alone publication available only on
the astro-ph archive; it will not be published separately in a journal.
\end{abstract}


\section{Introduction}

In 1996 the author published a catalog of parameters for globular
clusters in the Milky Way (Harris 1996).  This critical list
has been frequently
used since then, over four different updates that have been made
available through the Web.  The last of these, however, was
eight years ago, and since then a huge amount of observational
work on individual clusters \emph{and} on large collections of
clusters has been carried out.  A new edition at this stage
is well justified to take advantage of this rich new vein of
material.  

The 2010 edition can be found at

\noindent \emph{physwww.mcmaster.ca/$\sim$harris/Databases.html}

\noindent As before, the publication includes two files:  the compiled database,
and a much longer separate file giving the complete bibliography and 
explanation of how each quantity is derived.  In total, for this edition
material from almost 300 published papers since 2002 has
been added.

\begin{figure}
\plotone{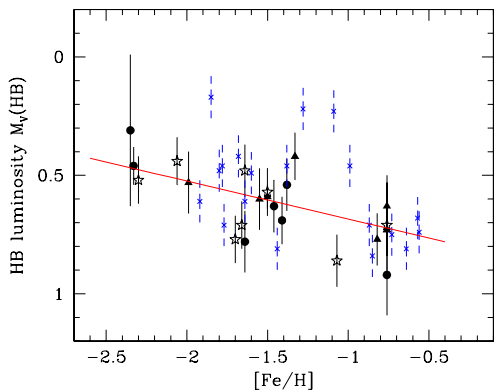}
\caption{Calibration of the horizontal-branch luminosity $M_V(HB)$ as a
function of cluster metallicity [Fe/H].  \emph{Solid dots} with errorbars
are values derived from dynamical parallax solutions for cluster distances
and RR Lyrae parallaxes;
\emph{Solid triangles} are values from main-sequence fitting to nearby subdwarfs;
\emph{Stars} are values either from eclipsing-binary distances or Fourier
analysis of RR Lyrae light curves.  Finally, \emph{blue crosses} with dashed errorbars
are HB levels from a homogeneous set of globular clusters in M31.
The fitted line (red) denotes the adopted solution $M_V(HB) = 0.16$[Fe/H] + 0.84.
}
\label{hb}
\end{figure}

\section{New Clusters, Coordinates, and Distances}

Seven new objects classified as globular clusters have been added
to the previous list, bringing the total to 157.  The new additions 
are Whiting 1, Koposov 1 and 2, FSR1735, BH261, and GLIMPSE-01 and 02.  
Several other candidates have been proposed in the literature over
the past decade, but as yet the evidence for those is less
convincing.  Many are equally likely to be ultra-faint
satellites of the Milky Way, or old open clusters in the Galactic disk.

       The fundamental list of coordinates has been completely reconstructed
for this edition of the catalog, thanks to numerous papers over the past
decade that provide distinct improvements over older work.  At present the center coordinates
for the clusters are now uncertain to typically 1-2 arcseconds, and
better than $1''$ in a few cases that have attracted especially intensive
study.

        The working definition of the cluster center that I use here
is the center of the overall light distribution, as defined for example
by concentric-aperture photometry, ellipse contour fitting, or light
profile fitting.  In a few special cases now in the literature alternate definitions
have arisen, such as the ``center of gravity" from starcounts or the
point around which the velocity dispersion would be maximized at zero radius.
At the level of $1''$ or so, these different definitions may disagree.

	The primary distance indicator for each cluster continues to be
the empirically observed horizontal branch level $V(HB)$.   Wherever possible,
$V(HB)$ denotes the mean $V$ magnitude of the RR Lyrae stars; for blue-HB
clusters, it denotes the magnitude of the HB at the blue edge of the
RR Lyrae region, while for red-HB clusters it denotes the mean magnitude
of the red HB stars themselves.  

The absolute calibration of $V(HB)$ adopted
here is
\begin{equation}
M_V(HB) \, = \, 0.16 {\rm [Fe/H]} \, + \, 0.84
\end{equation}
which is slightly fainter (though at most 0.05 mag) than in previous
editions of the catalog.  This calibration is constructed from the most
direct currently available distance measurements, including
(a) main-sequence fitting to metal-poor field subdwarfs
(Harris 2000; Grundahl et al.~2002; Gratton 2003; Layden et al.~2005;
Bergbusch \& Stetson 2009); (b) trigonometric parallax for field
RR Lyrae stars (Gratton 1998; Feast et al. 2008);
(c) dynamical parallax from combined solutions to radial velocity
and proper motion dispersions (Rees 1996; McLaughlin et al.~2006;
van de Ven et al.~2006; van den Bosch et al.~2006);
(d) eclipsing binary solutions (Thompson et al.~2001, 2010); and
(e) Fourier analysis of light curves or Baade-Wesselink method
for RR Lyraes (Olech et al.~2001; Cacciari et al.~2005; Dekany
\& Kovacs 2009; Arellano Ferro et al.~2010; Zorotovic et al.~2010).

The results for these methods are shown in Figure 1.  A weighted
fit gives
\begin{equation}
M_V(HB) \, = \, (0.165 \pm 0.045) {\rm [Fe/H]} \, + \, (0.863 \pm 0.071).
\end{equation}

A useful supplement to these Milky Way results can be obtained from the
observed HB levels in M31 globular clusters if we adopt a fiducial distance
for M31.  Rich et al.~(2005) provide a set of homogeneous, HST-based 
photometry and metallicities for 19 such clusters.  These are also shown
in Figure 1, for an adopted distance modulus $(m-M)_0 = 24.47$ from 
a combination of several standard candles including Cepheids, RR Lyraes,
Mira variables, and planetary nebulae.  A weighted fit for all 39 results,
both M31 and Milky Way, gives
\begin{equation}
M_V(HB) \, = \, (0.160 \pm 0.033) {\rm [Fe/H]} \, + \, (0.844 \pm 0.049)
\end{equation}
which is highly consistent with the trend from the Milky Way alone.  This second
version is essentially the calibration adopted for the catalog.
The rms scatter around these best-fit relations is $\pm 0.11$ magnitude.

\section{Metallicities and Velocities}

Considerable new work on abundances particularly from high dispersion spectroscopy
has been accomplished for individual clusters in the past decade.
The new heavy-element abundance scale adopted for the catalog is
the one established by Carretta et al.~(2009).
This represents a fundamental shift from the older Zinn
\& West (1984) metalllicity scale used in previous editions
of the catalog and by most other writers before the past decade.
The arguments for making this change of base are well
detailed by Carretta et al.~(2009 and earlier papers), but
essentially reduce to the fact that the traditional Zinn/West
scale was calibrated against only a handful of high-dispersion
spectroscopic [Fe/H] values available at that time.  In the
subsequent $\sim 30$ years, far more high-dispersion, high signal-to-noise
spectroscopic measures for vastly more clusters have been produced,
along with superior abundance analysis methods based on more
advanced model atmospheres.

        Carretta et al.~(2009) publish mean [Fe/H] values for
95 clusters derived from a weighted average of
(a) their own high-dispersion spectra in that paper and in
earlier papers from their group,
(b) the Kraft \& Ivans (2003) spectroscopic
metallicities transformed to their [Fe/H] scale,
(c) the Zinn \& West (1984) Q39 photometric
metallicity index transformed to their [Fe/H] scale, and
(d) the Rutledge et al.~(1997) $W'$ calcium
index transformed to their [Fe/H] scale.

These 95 transformed and averaged values 
are used as the homogenous, modern basis for the present catalog listing.
In addition to this basic list, dozens of additional
spectroscopic values of [Fe/H] for many individual clusters
are averaged in from all other studies not already included
in the sources listed above. Lastly, metallicities based on photometry (rather than
spectroscopy) are used for several clusters in which no other
estimates can be found.  These are mostly the clusters in the
Galactic center region that are very heavily reddened and where
every kind of observation is more difficult.  The photometric metallicities
are in most cases constructed from recent color-magnitude analyses 
and use parameters such as the
slope and color of the red-giant branch in various optical and
near-infrared bandpasses.

Similarly many new measurements of cluster radial velocities have 
been incorporated, though these have not required any fundamental
shift of basis.  Almost all the clusters in the Milky Way now have
useful, contemporary measurements of both metallicity and velocity.

\begin{figure}
\plotone{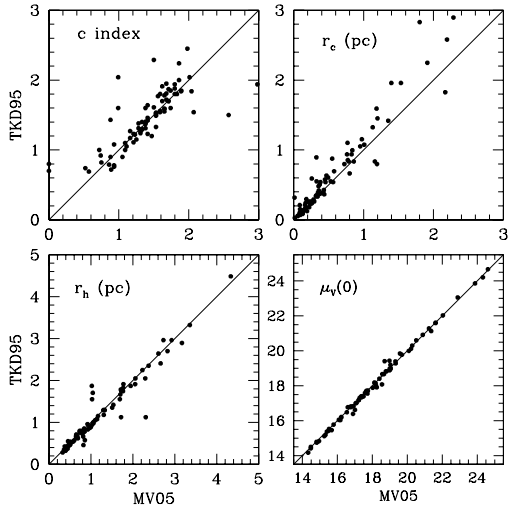}
\caption{Correlations of King-model structural parameters from two
recent studies, McLaughlin \& van der Marel (2005) and 
Trager, King \& Djorgovski (1995).  Successive panels show
Trager et al. vs. McLaughlin \& van der Marel for the central
concentration $c$, core radius $r_c$, half-light radius $r_h$,
and central surface brightness $\mu_V(0)$.
}
\label{king}
\end{figure}

An entirely new addition for this edition is the internal velocity
dispersion $\sigma(v_r)$, a critically important quantity for independent
calibration of cluster mass as well as all dynamical studies.  
The major starting point for this parameter continues to
be the work of Pryor \& Meylan (1993), who collected and analyzed the available
measurements up to that point.  Numerous additional papers are now to be
found with useful $\sigma(v_r)$ data, though in most cases these are 
only new measurements for the same clusters
as in the Pryor/Meylan compilations.   Thus unfortunately more than half the clusters do not yet
have a direct measurement of their internal velocity dispersion.
The velocity dispersion as quoted in the catalog is the
estimated central value (at or near $r=0$), sometimes relying on small
model extrapolations from velocity or proper motion data within the
cluster core, or in the best cases from the complete velocity dispersion profile.

\section{Structural Parameters}

Structural parameters for globular clusters continue to be useful for
a wide range of purposes such as discussions of dynamical relaxation,
core collapse, formation of exotic objects (black holes, degenerate binaries,
blue stragglers and so forth), or evolution of clusters in the tidal
field of the galaxy.  The basic measureables include core radius $r_c$,
half-light or effective radius $r_h$, central concentration $c$ as defined
in the King-model sense as the ratio of tidal to core radius, and
central surface brightness at zero radius $\mu_V(0)$.  

The starting points for the present list of 
structural parameters are the major compilations
of McLaughlin \& van der Marel (2005) and Trager et al.~(1993, 1995).
MV05 use the same raw data as TKD95, and direct comparison of
their derived values from King (1966) dynamical profile models
shows good internal agreement except for a few cases where the
profiles are unusually complex or low signal-to-noise.  The direct
comparisons are shown graphically in Figure 2.  In addition to these
studies, however, almost two dozen other recent sources are used
for individual clusters.

The ``tidal'' or limiting radius commonly labelled $r_t$ is, however,
no longer listed specifically in the catalog.  Recent work (see
particularly McLaughlin \& van der Marel) has made it clear that
$r_t$ is a much more model-dependent number than the other
parameters, because it requires an extrapolation of the observed
starcounts versus radius to the point at which the stellar density
drops to the (unmeasureable) level of zero.  It can also be
affected by the existence of tidal tails or envelopes of gradually
escaping stars outside the nominal tidal boundary, which are
problematic for the simple structural models.

\section{Summary}

The raw census of Milky Way globular clusters continues to increase,
albeit slowly.  Objects discovered and identified fairly convincingly
as globular clusters over the past $\sim 20$ years have almost invariably
been small, faint systems either deep in the heavily reddened bulge
region or in the remote halo.  A few more discoveries of such objects
can be expected from deep,
wide-field surveys covering the halo particularly.

Some of the more commonly used distribution functions drawn from
the new catalog are shown in Figures 4 and 5:  metallicity [Fe/H],
luminosity $M_V^t$, half-light radius $r_h$, and median
relaxation time $t(r_h)$.  The basic features of these distributions
have not changed in any major way, but they are now more confidently
filled out and include almost every known cluster.

More than anything else, this new edition of the catalog is a
testament to the rate of intensive observational work on a variety
of fronts during the past decade.  It is a pleasure to acknowledge
the ever-increasing quality and depth of such work on which any
compilation such as this is built.

\begin{figure}
\plotone{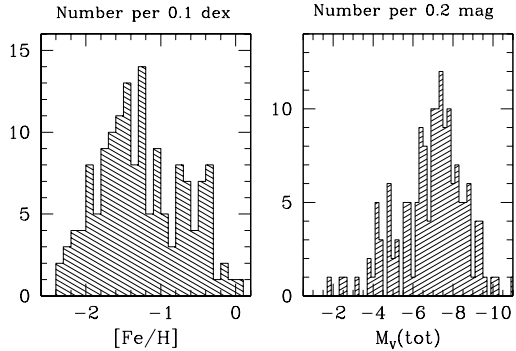}
\caption{\emph{Left panel:} Distribution of metallicities [Fe/H] for
all Milky Way globular clusters.  \emph{Right panel:} Distribution
of luminosities $M_V^t$.
}
\label{fehmv}
\end{figure}

\begin{figure}
\plotone{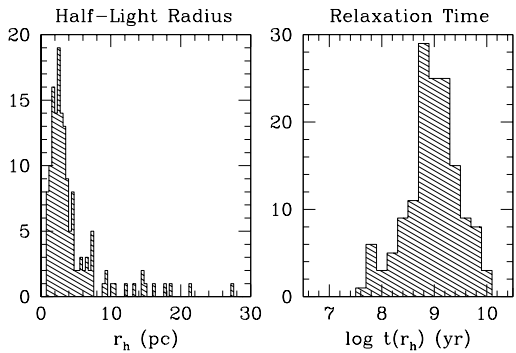}
\caption{\emph{Left panel:} Distribution of effective (half-light) radii for
all Milky Way globular clusters.  \emph{Right panel:} Distribution
of median internal relaxation times $t(r_h)$.
}
\label{rhtimes}
\end{figure}

\acknowledgments
This work was supported by the Natural Sciences and Engineering
Research Council of Canada through research grants to the author.
I am grateful for the hospitality at ESO (Garching) and at Mount
Stromlo Observatory where most of this work was done.   

%***************************************************************************
%*************************REFERENCES*************************************
%***************************************************************************

\end{document}